\documentclass[12pt]{article}
\usepackage{epsf}
\usepackage{cite}
\usepackage{amsfonts}
\def\calo{{\cal O}}

\begin{document}
\bigskip
\hspace*{\fill}
\vbox{\baselineskip12pt \hbox{hep-th/0002111}}
\bigskip\bigskip\bigskip

\centerline{\Large \bf On geodesic propagators and black hole holography
}
\bigskip\bigskip\bigskip
\renewcommand{\thefootnote}{\fnsymbol{footnote}} 

\centerline{\large Jorma Louko${}^1$\footnote{\tt
Jorma.Louko@nottingham.ac.uk}, 
Donald Marolf${}^2$\footnote{\tt
marolf@suhep.phy.syr.edu} and Simon F. Ross${}^3$\footnote{\tt 
S.F.Ross@durham.ac.uk}}
\medskip
\centerline{${}^1$ School of Mathematical Sciences, 
University of Nottingham}
\centerline{University Park, Nottingham NG7 2RD, 
United Kingdom}
\vskip .5cm 
\centerline{${}^2$ Department of Physics, Syracuse University}
\centerline{Syracuse, NY 13244-1130, USA}
\vskip .5cm 
\centerline{${}^3$ Department of Mathematical Sciences, 
University of Durham}
\centerline{Science Laboratories, South Road, Durham DH1 3DT, 
United Kingdom}
\vskip 2cm

\begin{abstract}
One of the most challenging technical aspects of the dualities between
string theory on anti-de Sitter spaces and conformal field theories is
understanding how location in the interior of spacetime is represented
in the field theory. It has recently been argued that the interior of
the spacetime can be directly probed by using intrinsically non-local
quantities in the field theory.  In addition, Balasubramanian and Ross
[hep-th/9906226] argued that when the spacetime described the
formation of an AdS$_3$ black hole, the propagator in the field theory
probed the whole spacetime, including the region behind the horizon.
We use the same approach to study the propagator for the BTZ black
hole and a black hole solution with a single exterior region, and show
that it reproduces the propagator associated with the natural vacuum
states on these spacetimes. We compare our result with a toy model of
the CFT for the single-exterior black hole, finding remarkable
agreement. The spacetimes studied in this work are analytic, which
makes them quite special.  We also discuss the interpretation of this
propagator in more general spacetimes, shedding light on certain
issues involving causality, black hole horizons, and products of local
operators on the boundary.
\end{abstract}

\newpage
\setcounter{footnote}{0} 
\renewcommand{\thefootnote}{\arabic{footnote}} 

\section{Introduction}

The proposed duality between string theory on anti-de Sitter space and
lower-dimensional conformal field theory~\cite{juanads} provides a
non-perturbative definition of string theory, and could thus, subject
to the restriction on the asymptotic boundary conditions, cast a
bright light on many dark corners of quantum gravity. In particular,
the field theory description encompasses arbitrary fluctuations of the
metric and other fields in the interior, and should provide a fully
quantum description of the formation and evaporation of a black hole.
One of the major barriers to studying conceptual questions in quantum
gravity using this theory is our poor understanding of how an
approximately local classical (or semi-classical) spacetime
description of the physics emerges from the fundamental gauge theory
description, and the consequent absence of any intuition about how
this approximate locality breaks down under extreme conditions. (A
related problem is that in the regime where a classical spacetime
description is a good approximation, we don't have any other
quantitative description; see \cite{kabat:approx} for a recent attempt
to construct calculationally useful approximations.)

The connection between asymptotic behavior of the spacetime fields and
the field theory was one of the first subjects of
study~\cite{witten,gkp}, and it was subsequently shown that the map
between states in the field theory and states in spacetime identifies
the asymptotic behavior of the fields with the expectation values of
local operators in the field theory~\cite{bklt}.  This was used to
show a ``scale-radius duality'' for a variety of bulk sources, and for
wavepackets of supergravity fields -- the radial position of a bulk
probe is encoded in the scale size of the dual expectation values.
Dynamical sources for supergravity fields were studied in~\cite{dkk},
where the radial position of a source particle following a bulk
geodesic was reflected in the size and shape of an expectation-value
bubble in the CFT.  The expectation values of the operators produced
by spacetime sources were further studied in
\cite{bdhm,esko1,amanda,garysunny,bakrey,sumit1,joeetal}.

However, the simple scale-radius relationship seen in these studies is
a consequence of an isometry in pure AdS space which is dual to a
scale transformation in the conformal field theory, under which the
vacuum remains invariant.  For situations describing black holes,
which break the symmetries, the relationship between bulk position and
boundary observables will be more complicated \cite{dkk,cons:glue}.
The same phenomenon is apparent in the collision of two massless
particles to form a black hole in~\cite{joeetal}; after the particles
collide, their radial position is fixed, but the scales in the
boundary expectation values continue to evolve.

Furthermore, the asymptotic values of the fields are not sufficient to
reproduce the whole spacetime.  Since asymptotic values of fields in
AdS space are dual to the expectation values of local operators in the
CFT, it follows that such expectation values describe only a small
piece of the physical information.  A number of authors have studied
spacetime sources which do not change the asymptotic values of the
fields, such as particles in AdS$_3$ and spherical shells, and found
that the location of the shell or particles is encoded in non-local
operator expectation values in the field theory, such as the two-point
function and Wilson
loops~\cite{holopart,sfsg:shell,dan:shell,chep:shell}. Similar work is
described in~\cite{suss:loop}. Thus, non-local operators must be
included in any understanding of the bulk-boundary connection. A
particularly striking case is asymptotically AdS$_3$ spaces, where we
can describe a wide range of dynamics without relying on perturbations
around some background~\cite{matschull,holst-matschull}, and the
asymptotic metric only encodes the total mass and angular momentum of
the system.

In~\cite{holopart}, Balasubramanian and Ross used a stationary phase
approximation to obtain predictions for the propagator in the gauge
theory from the geodesics of supergravity solutions in which a black
hole was formed.  This propagator appeared to be sensitive to events
in the interior of the black hole.  Now, while the CFT may well encode
information about the black hole interior, the particular CFT
propagator studied in \cite{holopart} is in fact the restriction to
the boundary of AdS space of a propagator associated with the bulk
quantum field theory.  This raises certain issues about
causality\footnote{These issues were brought to our attention by Lenny
Susskind through his comments at the Val Morin workshop on Black
Holes, June 1999.} which we wish to clarify in the work below.  Some
general arguments are presented in section \ref{gen}.  In short, we
argue that the propagator studied in \cite{holopart} is in fact a
causal object, but that the stationary phase approximation is valid
only in appropriately analytic spacetimes and not in the actual
spacetime considered in~\cite{holopart}. However, even without the
stationary-phase approximation, the path-integral definition of the
propagator used in \cite{holopart} should generally lead to a result
which depends on the region inside the black hole; we argue that this
should be interpreted as an object which is defined by a mixture of
past and future boundary conditions.

We then proceed to explore the propagator in two analytic spacetimes
in order to see more precisely what sort of object it represents.  The
spacetimes that we consider contain black holes, but are static
outside the Killing horizon.  In those cases, the stationary phase
approximation is expected to be valid, and a computation of the
propagator reduces to a study of various geodesics in the bulk
spacetime.  We show that, in such cases, the propagator of
\cite{holopart} is in fact the boundary limit of a time-ordered
expectation value of a product of local bulk fields.  Our spacetimes
are the spinless BTZ black hole
\cite{BTZ} and the associated $\mathbb R \mathbb P ^2$
geon~\cite{louko:geon}.  We find that the propagator in each case
is associated with a natural vacuum state for linearized quantum
fields on the spacetime, and that geodesics passing behind the black
hole horizon play an important role in determining the structure of
this state.  The states are analogues of the Hartle-Hawking state,
and are defined by boundary conditions at past and future
infinity. The propagators in these cases are known to be Green's
functions of a (causal) wave equation, and sensitivity to `events'
behind the event horizon would once again seem to contradict this
causality. In this case, the resolution is that the analyticity of
these spacetimes implies that much of the information about the region
inside the event horizon is in fact `stored outside'. Note however
that knowledge of the region outside the Killing horizon is not enough
to determine what happens inside the (future) event horizon; we also
need access to the `white hole' region, inside the past event horizon.

The next section is devoted to a short commentary on the AdS/CFT
correspondence and to general arguments concerning the nature of the
calculations in \cite{holopart}.  Section \ref{geo} then reviews the
BTZ and geon spacetimes and determines the propagators on these
spacetimes given by the path integral of \cite{holopart}.  In section
4, these calculations are compared to the propagator in the dual
CFT. We discuss the extension of the propagator calculation to the
rotating BTZ black hole spacetime in an appendix.

\section{The setting and the approximations}
\label{gen}

We use this section to set the stage for our later calculations.  The
most relevant elements of the AdS/CFT correspondence are briefly
reviewed in section \ref{bcl}.  This allows us to discuss the
particular regime in which we use the correspondence and to comment on
certain subtleties.  We then address the stationary phase
approximation and the issue of causality in section \ref{state}.
Section \ref{interp} includes a few further comments on the
interpretation of the propagator.

\subsection{The correspondence in the bulk classical limit}
\label{bcl}

While the AdS/CFT correspondence is conjectured to relate the full
quantum theories associated with bulk string theory and the CFT, it is
fair to say that this correspondence is best understood in the
neighborhood of the vacuum.  In that region, a useful way to describe
the correspondence is in terms of the partition functions $Z_{CFT}$
and $Z_{bulk}$, which in both cases are functions of external sources
that may be coupled to the CFT and to the boundary of the AdS space.
Recall that the CFT lives on a spacetime which may be identified with
the boundary of AdS$_3$.  The partition functions are equal and, by
differentiating them, we may arrive at relations between propagators
and correlators in the two theories.  For example, differentiating
twice yields the relation~\cite{bdhm}
\begin{equation} \label{prop1}
\langle \calo_\partial (b), \calo_\partial (b') \rangle_{\partial} =
\lim_{\epsilon \rightarrow 0} \epsilon^{-2 \Delta} \langle \calo_{B}
(b_\epsilon) \calo_B (b'{}_\epsilon) \rangle_B
\end{equation}
between the propagators in the boundary and bulk, where the bulk
operators $\calo_B$ are at points $b_\epsilon$, $b'{}_\epsilon$ in the
bulk that approach the points $b,b'$ in a certain way as $\epsilon
\rightarrow 0$.  (also see~\cite{witten,gkp}).  This is a relation
between the Euclidean propagators or, via analytic continuation,
between the Feynman propagators in the respective vacuum states.
Since we are in the vacuum state, operators on the right-hand
side may be viewed as fields on AdS space.

In the work below, we again wish to consider a propagator or
correlator.  However, we wish to work in a regime that is rather far
from the vacuum state.  We consider a state in which the bulk string
theory is nearly classical and contains, or is in the process of
forming, a large black hole.  Since the bulk string theory is nearly
classical, quantum fluctuations are infinitesimal and are well
approximated by linear fields.  In terms of the CFT, this is the limit
of large 't Hooft coupling.  While this is not the classical limit of
the CFT, it is a limit in which we again expect certain kinds of
classical behavior (such as factorization of correlation functions
with infinitesimal corrections).

Now, by acting on the vacuum with a sufficient set of local operators,
we should be able to reach any state in the Hilbert space.  Thus, the
relation between the partition functions implies that any state
$|\Phi\rangle_\partial$ in the CFT will be associated with some state
$|\Phi\rangle_B$ in the bulk.  Unfortunately,  it is
difficult to describe this relationship in detail.  Nonetheless, given
any bulk state and an associated state in the CFT, it follows that
correlation functions in the CFT state are given much as above by the
limit of correlation functions in the bulk as the points are moved to
the boundary of the spacetime.

As stated above, the regime of interest here is the limit in which the
bulk spacetime is nearly classical and in which the quantum
fluctuations are effectively linear.  This is just the usual setting
of (free) quantum field theory in curved spacetime.  As a result, it
is clear that a given classical geometry does not determine a unique
quantum state, but rather determines an entire space of states for the
linearized fluctuations.  For globally static spacetimes, one can
identify a preferred vacuum state, though this is not generally
possible.  For example, in the familiar asymptotically flat black hole
spacetimes, the `natural' choices of state for the linear quantum
fields include the Hartle-Hawking vacuum as well as the Unruh vacuum,
and more complicated choices of state are possible as well.  The
particular choice of quantum state may be associated with initial
and/or final conditions satisfied by the linearized fluctuations.

Now \cite{holopart} used the relation (\ref{prop1}) to link a CFT
object to a bulk propagator.  As a result, some particular choice of
state, or perhaps several states or a class of states, for the
linearized bulk quantum fields must have been made implicitly.  We
note that in \cite{holopart} it was explicitly assumed that the
`propagator' for a scalar field $\phi$ in the bulk was given by the
path integral expression
\begin{equation}  
\label{picor}
\langle \phi (x) \phi(x') \rangle_{\rm FPI} = \int d{\cal P} e^{i
\Delta L({\cal P})},
\end{equation}
where $L({\cal P})$ denotes the length of the path ${\cal P}$.  
The measure $d{\cal P}$ was not specified in detail as
the intention of \cite{holopart} was to use the expression
(\ref{picor}) only in the semiclassical approximation.  The
subscript FPI reminds us that this is the object defined by a Feynman
path integral, to distinguish it from other two-point functions that
we may wish to discuss.  The conventions are set here so that
spacelike paths have positive imaginary length, while timelike paths
have real length.  The question we wish to explore is whether this is
in fact the 2-point function of any bulk quantum state and, if so,
just which state it represents.

Now, the two-point function alone does not uniquely determine the
quantum state.  However, for linear fields there is the notion of a
quasi-free state (see, e.g., \cite{RMW}), also known as a Gaussian
state, in which the higher connected n-point functions vanish, and all
of the structure is in fact determined by the two-point function.  It
is therefore natural to attempt to associate the calculations of
\cite{holopart} with a quasi-free state of the linearized bulk fields.
We will show below that, on the BTZ black hole spacetime, the
expression (\ref{picor}) does in fact yield the 2-point function of
the Hartle-Hawking vacuum state.  Similarly, on the $\mathbb R \mathbb
P^2$ geon spacetime, it yields the 2-point function of the so called
geon-vacuum, the analogue of the state discussed in \cite{LM} for
asymptotically flat geons.  These states are in fact quasi-free.  In
subsection \ref{interp} below, we will discuss to what extent we can
draw the same conclusion in more general spacetimes.

\subsection{Causality and the stationary phase approximation}
\label{state}

After calculating the propagator (\ref{picor}) using a stationary
phase method, it was found in \cite{holopart} that this propagator was
sensitive to events happening behind a black hole's event horizon.
This raises certain issues about causality.  Stated most simply, we
have noted (see eq. \ref{prop1}) that the correlation functions in the
CFT are (up to a rescaling) the boundary limits of correlation
functions in the bulk.  However, in the current context of bulk
correlators for linear quantum field theory in curved spacetime, it is
well known that the evolution is causal.  An operator at any point in
the spacetime can be expressed purely in terms of operators in its
past light cone.  How, therefore, are we to interpret the results of
\cite{holopart} which suggest that correlation functions of such
operators near the boundary are sensitive to the interior of the black
hole?

In order to address this question, we first provide a few words on the
general interpretation of the propagator (\ref{picor}).  Let us first
note that there are at least two natural ways that we might try and
interpret this object.  The first is as a (time-ordered) correlation
function in some quantum state.  For definiteness, let us use the word
`state' in the sense of algebraic quantum field theory.  This means
that a `state' $\rho$ may be either a pure state or a mixed state and
that we would try to interpret (\ref{picor}) as $Tr \big(\rho
T(\phi(x) \phi(y)) \big)$ for some $\rho$.  The second natural choice
is to try to interpret the propagator as the time-ordered version of a
transition amplitude: $\langle \alpha |T(\phi(x) \phi(y) )| \beta
\rangle$.  In either case, however, the propagator would be a Green's
function for the wave operator and thus a causal object.

Thus, we need to know whether the propagator (\ref{picor}) does in
fact yield a Green's function for the wave operator $\nabla^2$.  That
this is the case may be argued as follows.  Let us consider the
spacetime as the configuration space of a ``non-relativistic
particle'' and take $H = \nabla^2$ to be its
Hamiltonian\footnote{Which, in this case, is unbounded from below due
  to the Lorentzian signature of the spacetime.}.  As usual, we may
write
\begin{equation}
\frac{-i}{H} = \int_0^\infty e^{- iN(H - i \epsilon)} dN,
\end{equation}
so that the object on the right hand side defines a
Green's function for the wave operator.  By the usual path
integral skeletonization
arguments, one can write this as
\begin{equation} 
\label{skel}
\langle x| \frac{-i}{H}| y \rangle = \int_0^\infty  dN
\int Dx Dp \exp{i \int_0^1 [\dot{x}p - N(p^2 + m^2)]d\lambda}, 
\end{equation}
where $\dot{\phantom{x}} = d/d\lambda$.  We will see in a moment that
(\ref{skel}) is just the path integral (\ref{picor}) in another
form. Alternatively, (\ref{skel}) could be taken as the definition of
the path integral (\ref{picor})~\cite{teitel-pi1,henn-teitel,piINST}.

The path integral above contains the action for a free relativistic
particle.  Note, however, that while such particles are typically
associated with a time reparametrization invariance, there is no such
explicit invariance above.  We may thus consider (\ref{skel}) to be a
gauge-fixed path integral, using in particular the gauge $\dot{N}=0$:
\begin{equation}
\label{skel2}
\langle x| \frac{-i}{H}| y \rangle = \int_0^\infty  DN
\int Dx Dp \ \delta(\dot N) \exp{i \int_0^1 [\dot{x}p - N(p^2 + m^2)]
d\lambda}.
\end{equation}
The argument below will be more transparent if we change the gauge
fixing scheme to use a gauge condition that depends only
on the path $x(\lambda)$ through position space\footnote{A
complete such gauge fixing cannot be a smooth function of the
path $x(\lambda)$, but this need not concern us here.}.  Thus, 
we write
\begin{equation}
\label{skel3}
\langle x| \frac{-i}{H}| y \rangle = \int_0^\infty  DN
\int Dx Dp \  \Delta(x) \exp{i \int_0^1 [\dot{x}p - N(p^2 + m^2)]
d\lambda}
\end{equation}
where $\Delta(x)$ contains both the gauge fixing condition
and the associated Faddeev-Popov determinant.  Note that $\Delta(x)$
will depend only on~$x(\lambda)$.

Now, to lowest order in the WKB approximation, performing an integral
over some variable is equivalent to solving the associated classical
equation of motion and inserting the result back into the action.
Thus, we can do the integrals over $N$ and $p$ and write the result as
follows:
\begin{equation}
\label{skel4}
\langle x| \frac{-i}{H}| y \rangle = 
\int Dx \Delta'(x) \exp{i L(x(\lambda))},
\end{equation}
where $L(x(\lambda))$ is the length of the path $x(\lambda)$
with exactly the same conventions as in (\ref{picor}).

The factor $\Delta'(x)$ denotes $\Delta(x)$ together with the various
path-dependent measure factors arising from the corrections to the WKB
approximation in integrating over $N$ and $p$.  Identifying $d{\cal P} =
\Delta'(x) Dx$, we find that our Green's function is just the
propagator (\ref{picor}).  Note that solving the equation of motion
for $N$ involves taking a square root.  For the timelike segments of
path the restriction $N >0$ was used to choose the appropriate branch.
For the spacelike segments, the appropriate branch is determined by
the details of the measure as discussed in~\cite{piINST}.  Note that
the action is an analytic function of both $N$ and $p$ so that we
expect no problems with the use of stationary phase methods here.
Thus, the propagator (\ref{picor}) is indeed a Green's function for a
wave operator.  That (\ref{picor}) satisfies Dirichlet boundary
conditions on the smooth part of the boundary at infinity can be seen
from the arguments of \cite{Fuller}.

At this point, we can now reduce our physical question about causality
in the setting of \cite{holopart} to a mathematical question about
solutions of the wave equation.  In \cite{holopart} further stationary
phase methods were used to argue that, to leading order, the
propagator was in fact determined by the shortest geodesic connecting
the points $x$ and $y$.  The authors considered a spacetime that was pure
AdS before a certain spacelike hypersurface $\Sigma$ on which two
massless point particles entered through the boundary at infinity.
{}From this it is clear that two points sufficiently far in the past
of $\Sigma$ can only be connected by geodesics that lie in the pure
AdS part of the spacetime.  Thus, the geodesic approximation leads to
the conclusion that, to the past of some hypersurface $\Sigma'$, the
propagator is just as it would be in pure AdS space\footnote{In that
  case, as we will discuss below, it is known to be the time ordered
  2-point function in the AdS vacuum.}.

Nonetheless, at sufficiently late times, it was shown in
\cite{holopart} that there are points outside the black hole such that
the shortest geodesic connecting them runs through the interior of the
black hole.  It was therefore concluded that the propagator
(\ref{picor}) outside the black hole was sensitive to events occurring
inside the black hole.

In order to eliminate certain technical worries, let us consider a
family of generalizations of the spacetime constructed in
\cite{holopart}.  Imagine replacing the singular null particles with a
distribution of null fluid of compact support.  Since there is no
local gravitational dynamics in 2+1 dimensions, the resulting
spacetime is easily made identical to that of \cite{holopart} outside
of the region occupied by the null fluid.  Until the formation of the
black hole singularity, the resulting spacetime is then
smooth\footnote{It is not, however, asymptotically AdS where the null
  fluid enters the spacetime.  We shall assume that this does not
  cause any further complications.}.  If the field $\phi$ for which we
compute the propagator does not couple to the null fluid, then the
definition of the propagator on this spacetime remains just
(\ref{picor}).  Thus, we have a complete specification of the
propagator, up to issues associated with the black hole
singularity\footnote{Such issues certainly exist.  For example, if we
  take (\ref{skel}) as the definition of (\ref{picor}), the black hole
  singularity will imply that $H$ is not essentially self-adjoint and
  that some particular self-adjoint extension should be chosen.  Here,
  we simply assume that some such choice has been made.}.

Suppose now that we arrange things such that the two bits of null
fluid actually collide inside the black hole.  That is, suppose that
at some event the supports of the two distributions of fluid overlap.
Note that, depending on the sort of null fluid used, various outcomes
are possible.  Some sorts of fluid would interpenetrate readily while
other sorts would bounce solidly off of each other.  The outcome
should affect some of the geodesics mentioned above that connect two
points near infinity by passing through the interior of the black
hole.

Now we see that we have a real contradiction at hand.  On the one
hand, we have the statement that the propagator at early times is the
AdS vacuum correlator -- independent of what goes on in the black hole
interior.  Also, we know that the propagator satisfies the wave
equation and so evolves in a causal fashion.  Thus, the propagator at
points outside the black hole can be expressed in terms of initial
data on an early hypersurface in a manner that is independent of what
goes on in the black hole interior.  Thus, the propagator outside the
black hole cannot in fact depend on events inside the black hole.
This is in direct contradiction to the conclusion of the previous
paragraph. 

The resolution seems to be that the geodesic `approximation' is not in
fact a valid approximation\footnote{It is also a logical possibility
  that the approximation is valid, but simply unstable in a manner
  that causes higher order effects at early times to evolve into lower
  order effects at late times.}.  In retrospect, it seems quite likely
that this approximation fails for such a spacetime.  Note that to
arrive at the geodesic approximation, one would use a stationary phase
argument to solve the classical equations of motion corresponding to
the action $m \int \sqrt{-\dot{x}^2}$.  While the stationary point
(the spacelike geodesic) does indeed lie on the original contour of
integration (real values of $x$), this contour is not a steepest
descent contour through the stationary point.  In particular, in a
Lorentzian signature spacetime, a spacelike geodesic is not a path of
minimal length.  As a result, if one wishes to argue that the
stationary point dominates, one must first analytically continue the
action to complex values of the coordinates and attempt to deform the
original contour to the contour of steepest descent.

Now, the action involves the metric $g_{ab}(x)$.  To avoid the issue
of the singularities, let us consider the smoothed spacetimes with
null fluid sources.  Since the fluid density vanishes in an open
region, but not in the entire spacetime, it is clear that such
spacetimes are not analytic and that continuation is problematic.
Thus, it is not at all clear that steepest descent methods should
succeed in this case, and we are happy to associate their failure with
nonanalyticities of the spacetime.

While this seems to settle the issue nicely, we should mention for
completeness that, if one excises the region of non-zero fluid density
from the spacetime, the resulting spacetime does have a real analytic
atlas and can be continued.  Presumably, excising the region occupied
by the fluid prevents one from deforming the contours as one would
like.

\subsection{Interpreting the Propagator}
\label{interp}

Having ruled out the use of the geodesic approximation in general,
what are we to conclude about the full propagator (\ref{picor})?  In
principle, picking any two points $x$ and $y$ in the spacetime, the
path integral includes contributions from paths connecting them that
explore arbitrarily far into the future.  As a result, even in the
spacetime studied in \cite{holopart}, it is far from clear that the
propagator at early times is independent of events in the interior of
the black hole.  It seems likely that the propagator does not
correspond to a fixed initial condition, but instead to some mixture
of initial and final conditions.  In this case, the propagator between
points near the boundary at late times may depend on events inside the
black hole as well. That is, it may be possible to choose a state or
states in such a way that the two-point function reproduces the
important qualitative features found in the calculation in
\cite{holopart}. Such a state (or states) will involve a mixture of
initial and final conditions, reflecting the fact that the form of the
two-point function in \cite{holopart} depends on the formation of the
black hole in the future.

Let us return to the two natural interpretations of the propagator
mentioned above: as a time ordered expectation value in some
quasi-free state, and as a time ordered transition amplitude between
two states.  We note that either is compatible with the above
observations.  In the case of the expectation value, it may simply be
the case that the quantum state itself is one that is naturally
defined by a combination of retarded and advanced boundary conditions,
and so is free to depend on events in the interior of the black hole.
We note that the Hartle-Hawking state for an asymptotically flat black
hole is an example of such a state that is naturally associated with
boundary conditions in both past and future, while the Unruh state is
associated only with boundary conditions in the past.  In the case of
the transition amplitude, both states may involve such `mixed'
boundary conditions, or perhaps one is defined by retarded boundary
conditions and one by advanced boundary conditions.

In spacetimes that are asymptotically flat at both timelike and
spacelike infinity, the propagator (\ref{picor}) can be shown to
define a transition amplitude \cite{DeWitt}.  On the other hand, the
work of Wald \cite{Wald2} effectively shows that (\ref{picor}) defines
an expectation value for globally static spacetimes (without
horizons). It is also known to give the expectation value of
time-ordered fields in the Hartle-Hawking state on the Kruskal
spacetime, though the status of this question on a general black hole
spacetime is not yet understood \cite{RMW}.  We will see that an
expectation value is once again obtained on the spinless BTZ spacetime
and the associated $\mathbb R \mathbb P^2$ geon.

\section{The geodesics in AdS$_3$ and quotient spacetimes}
\label{geo}

We have argued in section \ref{gen} that stationary phase methods do
not in general yield a valid approximation to the FPI propagator
(\ref{picor}).  Nevertheless, one may ask if there are cases in which
it does provide a valid approximation and, if so, whether geodesics
passing behind the horizon play any important role.  We shall see in
this section and the next that the answer to both of these questions
is in the affirmative.

In the present section, we consider the lengths of spacelike geodesics
in the AdS${}_3$, spinless BTZ, and ${\mathbb R} {\mathbb P}^2$ geon
spacetimes.  As these spacetimes are real Lorentzian sections of
holomorphic complex manifolds, one may expect the geodesic
approximation to succeed in these cases.  Indeed, it is known
\cite{holopart} to succeed in yielding the vacuum correlator on
AdS${}_3$. In the following section, we consider the propagators
obtained through this approximation, and compare to what we know about
the field theory. This will allow us to explicitly check the agreement
with certain CFT calculations and to trace the role of geodesics
passing through the interior of the black hole.  The final agreement
provides additional confirmation of the accuracy of the geodesic
approximation in these cases.

In fact, these calculations are not truly independent. Since the
spinless BTZ and ${\mathbb R}{\mathbb P}^2$ geon spacetimes are
quotients of AdS${}_3$, a method of images argument together with
analytic continuations and the uniqueness of the Euclidean Green's
functions shows that the success of the geodesic approximation to
(\ref{picor}) in reproducing the vacuum correlator on AdS$_3$ implies
that it must also approximate the Hartle-Hawking correlation function
for the spinless BTZ hole and the related geon correlation function
(see \cite{LM}) on the ${\mathbb R}{\mathbb P}^2$ geon. Thus, in
these cases the FPI propagator gives the expectation value of a
time-ordered product of fields in a quasi-free state.

\subsection{Geodesics of AdS$_3$}
\label{review}

The AdS$_3$ spacetime can be constructed as the hyperboloid 
\begin{equation}
(T^1)^2 + (T^2)^2 -(X^1)^2 - (X^2)^2  = 1
\end{equation}
in a flat embedding space with metric
\begin{equation}
ds^2 = -(dT^1)^2  -(dT^2)^2 +(dX^1)^2 +(dX^2)^2. 
\end{equation}
Here, we are choosing units so that the AdS length scale $\ell$
(related to the cosmological constant) is one. A set of intrinsic
coordinates on AdS$_3$ is given in terms of these embedding
coordinates by 
\begin{equation}
T^1 = \cosh \chi \cos \tau, T^2 = \cosh \chi \sin \tau, X^1 = \sinh
\chi \sin \varphi, X^2 = \sinh \chi \cos \varphi,
\end{equation}
where $\varphi$ has period $2\pi$, and $0 \leq \chi \leq \infty$. For
the hyperboloid, $\tau$ is also periodic with period $2\pi$, but we
pass to the covering space, and take $\tau$ to run between
$\pm\infty$. In terms of these coordinates, the metric is \begin{equation} 
ds^2 = d\chi^2 + \sinh^2\chi \, d\varphi^2 - \cosh^2\chi \, d\tau^2 =
\left({2 \over 1- \rho^2}\right)^2 (d\rho^2 + \rho^2 d\varphi^2) -
\left( {1 + \rho^2 \over 1 - \rho^2} \right)^2 d\tau^2.
\label{adsmet} 
\end{equation} 
In the second equality, we have defined a new radial coordinate $\rho
= \tanh(\chi/2)$, so $0 \leq \rho \leq 1$. Fixed $\tau$ surfaces have
the Poincar\'e disc geometry, and the dual CFT is defined on a
cylinder isomorphic to the $\rho = 1$ boundary.

We will need the length of the unique geodesic traveling between
$(\tau,\chi_{{\rm m}},\pm\varphi_{{\rm m}})$. Now, since the metric at
fixed $\tau$ is that of the Poincar\'e disc, equal-time geodesics of
(\ref{adsmet}) are circle segments obeying the equation
\begin{equation} 
\tanh\chi \, \cos(\varphi - \alpha) = \cos(\beta),
\label{geodeq} 
\end{equation} 
where the geodesic reaches the $\chi = \infty$ boundary at $\varphi =
\alpha \pm \beta$. Setting $\alpha=0$, the unique geodesic between
the boundary points $(\tau,\pm \beta)$
intersects $\chi = \chi_m$ at $\varphi_m^\pm$ which are fixed by
\begin{equation} 
\tanh\chi_{{\rm m}} \, \cos\varphi_{{\rm m}}^{\pm} = \cos(\pm\beta),
\label{phimax} 
\end{equation} 
which implies that
\begin{equation} 
-\varphi_m^- = \varphi_m^+ \equiv \varphi_m.
\end{equation} 

Integrating (\ref{adsmet}) 
yields the length of the geodesic connecting $(\tau,\chi_{{\rm
m}},\pm\varphi_{{\rm m}})$:
\begin{equation} 
L(\varphi_{{\rm m}},-\varphi_{{\rm m}}) = 2 \ln\left[ \sinh\chi_{{\rm m}} \,
\sin\varphi_{{\rm m}} + (\sinh^2\chi_{{\rm m}} \, \sin^2\varphi_{{\rm
m}} + 1)^{1/2} \right]. \label{length1}
\end{equation} 

\subsection{Spacelike geodesics on the spinless BTZ hole}
\label{spin0BTZ}

The spinless BTZ hole is obtained by taking the quotient of the region 
$T^1 > |X^1|$ of AdS$_3$ by the isometry $\exp(2\pi r_+ \xi)$, where
$\xi$ is the Killing vector
\begin{equation} 
\label{quot}
\xi = 
X^1 \frac{\partial}{\partial T^1} 
+ 
T^1 \frac{\partial}{\partial X^1} 
\ \ . 
\end{equation}
To express this geometry in the Schwarzschild-like coordinates 
of the original papers~\cite{BTZ}, we introduce on the
region $X^2 > |T^2|$, $T^1 > 0$ of AdS$_3$
the coordinates $(t,r,\phi)$ by 
\begin{eqnarray} 
T^1 &=& {r \over r_+} \cosh(r_+ \phi),  \nonumber \\
X^1 &=& {r \over r_+} \sinh(r_+ \phi), \nonumber \\
T^2 &=& \left( {r^2 \over r_+^2} -1\right)^{1/2} \sinh(r_+ t ),
\nonumber \\ 
X^2 &=& \left( {r^2 \over r_+^2} - 1 \right)^{1/2} \cosh(r_+ t).
\label{nonrot-BTZcoords} 
\end{eqnarray} 
$t$ and $\phi$ take all real values, 
$r > r_+$, and the metric takes the form 
\begin{equation} 
\label{btz}
ds^2 = - N^2 \, dt^2 + r^2 \, d\phi^2 
+ {1 \over N^2} \, dr^2;~~~~ N^2
= r^2 - 8 GM
\ \ , 
\end{equation}
where $M = r_+^2/(8G)$. 
The identification by $\exp(2\pi r_+ \xi)$ amounts to 
$(t,r,\phi) \sim (t,r,\phi+2\pi)$, and with this
identification the coordinates $(t,r,\phi)$ 
cover one exterior region of the BTZ hole. 

We are interested in geodesics between two points, $x_1$ and~$x_2$, 
in the exterior region of the hole. 
We take the value of $r$ at both points to be the 
same. To parametrize the locations of the points, let $y_1$ and $y_2$
be two points in AdS$_3$, respectively at 
$(t_1,r,\phi_1)$ and 
$(t_2,r,\phi_2)$, 
and let $x_1$ (respectively $x_2$) be the
equivalence class of 
$y_1$ ($y_2$). 
We write 
$\Delta\phi = \phi_2 - \phi_1$ and 
$\Delta t = t_2 - t_1$, and we assume that 
$|\Delta \phi+ 2\pi n| > |\Delta t|$ for all
integers~$n$. 
For fixed 
$t_1$, $t_2$,
$\phi_1$, and $\phi_2$, it is then straightforward to show that for
sufficiently large $r$ there 
are countably many spacelike geodesics connecting 
$x_1$ and $x_2$ in the BTZ hole. 

To calculate the lengths of these geodesics, we exploit the
symmetries to argue that the geodesic distance between $y_1$ and 
$y_2$ in AdS$_3$
is a function only of the chordal distance $D$ 
in the embedding space,
\begin{eqnarray}
D &=& - (\Delta T^1)^2 - (\Delta T^2)^2 + (\Delta X^1)^2 + (\Delta
X^2)^2 \nonumber \\
&=& { 4r^2 \over r_+^2} \sinh^2 \left( {r_+ \Delta \phi \over 2} \right)
- 4 \left({ r^2 \over r_+^2} - 1 \right) \sinh^2 \left( {r_+ \Delta t
\over 2} \right). 
\end{eqnarray}
By considering a simple example of a spacelike geodesic, we can show
that the relation between chordal distance and proper length $L$ is
\begin{equation}
\sinh^2 (L/2) = {D \over 4}.
\end{equation}
It then follows from the 
quotient construction 
that the lengths, $L_n(x_1,x_2)$, of the 
geodesics connecting $x_1$ and~$x_2$ in the BTZ hole 
have the large $r$
expansion
\begin{equation}
\exp\left[L_n(x_1,x_2)\right]
= 
\frac{2 r^2}{r_+^2} 
\left\{
\cosh \left[r_+(\Delta \phi + 2\pi n ) \right] 
- \cosh(r_+\Delta t)
\vphantom{A^A_A}
\right\}
+ O\left( 1 \right) 
\ \ , 
\label{L-expr-BTZ}
\end{equation}
where $n\in{\mathbb Z}$.

\subsection{Spacelike geodesics on the $\mathbb R \mathbb P^2$ geon}
\label{subsec:geodesics-on-geon}

Recall \cite{louko:geon} that the $\mathbb R \mathbb P^2$ geon is
obtained by taking the quotient of the region $T^1 > |X^1|$ of AdS$_3$
by the isometry that is the composition of $J_1 : \exp(\pi r_+ \xi)$ and
the involution $J_2: (T^1, T^2, X^1, X^2) \mapsto (T^1, T^2, X^1, -
X^2)$.  The resulting spacetime is not orientable, but one can
construct a related orientable spacetime from the product of the BTZ
spacetime with $T^4$.  If the moduli of the $T^4$ are chosen so that
there is an orientation-reversing involution $J_4$ of the torus, then
one obtains an orientable spacetime by taking the quotient with
respect to $J_1 \circ J_2 \circ J_4$.

Now, let $y_1$ and $y_2$ be points on AdS$_3$ as above, 
respectively at
$(t_1,r,\phi_1)$ and 
$(t_2,r,\phi_2)$, and suppose that 
$|\Delta \phi+ 2\pi n| > |\Delta t|$ for all integers~$n$. 
Let $x_1$ and $x_2$ be two points in the exterior region of the
geon, such that $x_1$ (respectively $x_2$) is the equivalence class of
$y_1$~($y_2$).  
For sufficiently large~$r$, one class of spacelike
geodesics connecting $x_1$ and $x_2$ is then obtained precisely as for
the BTZ hole, with the result (\ref{L-expr-BTZ}) for their lengths.
The second class of geodesics arises from the AdS$_3$ geodesics
connecting $y_1$ to the points ${\tilde y}_{2;n}$, located at
\begin{eqnarray}
&&T^1 = (r/r_+) \, \cosh[r_+ (\phi_2 + \pi + 2\pi n )] 
\ \ , 
\nonumber
\\
&&X^1 = (r/r_+) \, \sinh[r_+ (\phi_2 + \pi + 2\pi n)] 
\ \ , 
\nonumber
\\
&&T^2 = \sqrt{(r/r_+)^2 -1} 
\, 
\sinh(r_+ t_2) 
\ \ , 
\nonumber
\\
&&X^2 = - \sqrt{(r/r_+)^2 -1} 
\, 
\cosh(r_+ t_2) 
\ \ , 
\label{xtildetwo-location}
\end{eqnarray}
where $n\in\mathbb Z$. 
As 
\begin{eqnarray}
D(y_1,{\tilde y}_{2;n}) 
&=& 
2(r/r_+)^2 
\left\{ 
\cosh\left[r_+(\phi_2 - \phi_1 + \pi + 2\pi n )\right] -1 
\vphantom{A^A_A}
\right\} 
\nonumber
\\
&&
+ 
2 \left[ (r/r_+)^2 -1 \right] 
\left\{ 
\cosh \left[ r_+ (t_2 + t_1) \right]  + 1 
\vphantom{A^A_A}
\right\}
\ \ , 
\label{Dtilde-geon} 
\end{eqnarray}
the lengths ${\tilde L}_n(x_1,x_2)$ 
of these geodesics have the large $r$ expansion 
\begin{equation}
\exp\left[{\tilde L}_n(x_1,x_2)\right]
= 
\frac{2 r^2}{r_+^2} 
\left\{
\cosh \left[r_+(\phi_2 - \phi_1 + \pi + 2\pi n ) \right] 
+ \cosh\left[r_+ (t_2 + t_1) \right]
\vphantom{A^A_A}
\right\}
+ O\left( 1 \right). 
\label{L-expr-geon}
\end{equation}
It is precisely this class of geodesics that pass through
the black hole interior.  We note that all such geodesics
are {\it longer\/} 
than the shortest geodesic connecting $x_1$
and $x_2$ through the exterior region.  Thus, at first sight
one might think that geodesics passing through the interior
cannot be relevant to leading order.  Nonetheless, we shall
see in section \ref{subsec:propag-in-geon} that they do provide
the leading contribution to the two-particle correlations in the
geon vacuum, and that (\ref{L-expr-geon}) reproduces
expectations based on the dual CFT.

\section{Matching to the CFT}

It turns out that, due to difficulties in performing the various mode
sums, there are few exact results for the bulk correlators in the
spinless BTZ Hartle-Hawking state and in the geon vacuum.  We will
therefore proceed by comparing the limiting behaviors of
(\ref{BTZ-geodapp-expansion}) and (\ref{L-expr-geon}) with
expectations based on toy models of the dual CFT.  We shall see that
the agreement is surprisingly good.  This supports both the accuracy
of the bulk geodesic approximation in these cases and the ability of
the toy models to capture much of the physics of the CFT.  We first
review the calculation showing that the geodesic approximation in
AdS$_3$ reproduces the vacuum propagator, and then show that the
asymptotic behavior of (\ref{L-expr-geon}) reproduces the expected
two-particle correlations in BTZ and the geon.

\subsection{The propagator in AdS$_3$}
\label{limits}

We will now review the calculation of the equal time correlation
functions in the dual field theory for the AdS$_3$ geometry using the
(bulk) WKB approximation. A scalar field of mass $m$ in a spacetime
which is asymptotically AdS$_3$ is dual to an operator $\calo$ of
conformal dimension $\Delta = 1 + \sqrt{1 + m^2}$. The fiducial metric
for the CFT on the cylinder is related to the induced metric obtained
from (\ref{adsmet}) by a diverging Weyl factor.  To relate operators
to expectation values, we need to regulate this behavior by cutting
off the spacetime at a boundary defined by
\begin{equation} 
\rho_{{\rm m}}(\tau,\varphi) = 1 - \epsilon(\tau,\varphi),~~~~~
\epsilon(\tau,\varphi) = \epsilon(\tau,-\varphi),
\label{regulate} 
\end{equation} 
where $\epsilon$ is some smooth function of the boundary coordinates.
The symmetry of $\epsilon$ under $\varphi \rightarrow -\varphi$ is
chosen for simplicity. For the calculations relating to the BTZ black
hole and geon, we will take the cutoff surface to be at constant $r$
in the BTZ coordinates. According to~\cite{bdhm}, the Feynman
propagator for $\calo$ in the dual CFT is obtained from the spacetime
propagator between the corresponding points on the cutoff boundary at
$\rho_{\rm m}$ (also see~\cite{witten,gkp}),
\begin{equation} \label{prop}
G_\partial((\tau,\varphi),(\tau',\varphi')) = \epsilon^{-2 \Delta}
G_B({\bf B},{\bf B'}),
\end{equation}
where ${\bf B} = (\tau,\varphi,\rho_{\rm m}(\tau,\varphi))$. We will
only need the propagator when $\tau=\tau'$.  For ${\bf B,B}'$ causally
unrelated, the Green's function $G_B({\bf B,B}')$ in the leading order
semi-classical approximation is given by a sum over geodesics:
\begin{equation} 
G({\bf B},{\bf B}') = \sum_g e^{-\Delta \, L_g({\bf B},{\bf B}')}.
\label{sum1} 
\end{equation} 
Here $L_g$ is the (real) geodesic length between the boundary 
points and only spacelike geodesics contribute since 
$\tau = \tau'$.
 
By rotational invariance, it is sufficient to perform the calculation
for $\varphi = - \varphi'$.  For the particular case $1 -
\epsilon(\tau,\varphi) = \tanh(\chi_m/2) = const$, the length of the
geodesic connecting ${\bf B}$ and ${\bf B}'$ is given by
(\ref{length1}).  In fact, the symmetry $\epsilon(\tau,\varphi) =
\epsilon(\tau,-\varphi)$ guarantees that a corresponding result holds
for any such symmetric choice of $\epsilon$.  So, to leading order in
$\epsilon$, the geodesic length between the points ${\bf B, B'}$ is
\begin{equation} 
L({\bf B},{\bf B'}) = 2 \ln\left(2\sin\varphi \over \epsilon \right).
\label{geodlen} 
\end{equation} 
The bulk propagator is thus
\begin{equation} 
G({\bf B},{\bf B'}) = \left(2 \sin\varphi \over
\epsilon \right)^{-2\Delta}
\label{eqt2} 
\end{equation} 
in the $\epsilon \rightarrow 0$ limit, where the boundary metric is
$ds^2 = (1/\epsilon(\tau,\varphi)^2) ( -d\tau^2 + d\varphi^2)$.  This
correctly reproduces the CFT two-point correlator of \cite{witten} for
$\Delta \tau=0$ and $\Delta \varphi = 2\varphi$, since the CFT is
defined on the Weyl rescaled cylinder with metric $ds^2 = -d\tau^2 +
d\varphi^2$.  

\subsection{The propagator in BTZ}
\label{prop-in-BTZ}

We now apply the bulk geodesic approximation method of \cite{holopart}
to the Green's function on the boundary of the spinless BTZ hole,
using the geodesic length (\ref{L-expr-BTZ}). The geodesic
approximation to the path integral (\ref{picor}) reads
\begin{eqnarray}
\langle \phi (x_1) \phi(x_2) \rangle_{\rm FPI} 
&&= \int d{\cal P} e^{i
\Delta L({\cal P})}
\approx 
\sum_n\exp\left[-\Delta L_n(x_1,x_2)\right]
\nonumber
\\
&&=
\left(\frac{r_+^2}{2 r^2} \right)^\Delta
\sum_{n=-\infty}^{\infty} 
\frac{1}
{
\left\{
\cosh \left[r_+(\Delta \phi+ 2\pi n) \right] - 
\cosh(r_+\Delta t )
\vphantom{A^A_A}
\right\}^\Delta
}
\nonumber
\\
&&
\phantom{xx}
+ 
O\left( \left( \frac{r_+^2}
{r^2} \right)^{\Delta+1}
\right) 
\ \ . 
\label{BTZ-geodapp-expansion} 
\end{eqnarray}
This is the bulk propagator; to relate it to the boundary propagator,
we observe that at large $r$, the boundary metric in the BTZ spacetime
is $ds^2 = r^2 (-dt^2 + d\phi^2)$. Since we want the CFT to live on
the same cylinder as above, the boundary propagator is given by
$G_\partial= r^{2 \Delta} G_B$ in this case. The rescaling thus
precisely cancels the $r^{-2\Delta}$ in the prefactor. 

We note that this Green's function is manifestly periodic in the
global time $t$ in the imaginary direction, and the period $2\pi/r_+$
is the inverse of the spinless BTZ temperature.  It is thus plausibly
identified with the propagator in the analogue of the Hartle-Hawking
state for this black hole. We also note that the geodesics used in
this calculation lie entirely outside the black hole. This calculation
successfully reproduces the result given in \cite{esko1} for the
spinless BTZ black hole. A similar agreement is obtained for the
rotating BTZ black hole in appendix A.

\subsection{The propagator in the single-exterior black hole}
\label{subsec:propag-in-geon}

We now proceed to address the dual CFT propagator associated with the
bulk FPI propagator $\langle x_1 \ x_2 \rangle_{\rm geon}$ on the
${\mathbb R} {\mathbb P}^2$ geon.  Now, the corresponding path
integral can be written as a sum of two contributions:
\begin{equation}
\langle x_1 \ x_2 \rangle_{\rm geon} = \langle x_1 \ x_2 \rangle_{\rm
BTZ}
+ \langle x_1 \ J(x_2) \rangle_{\rm BTZ}
\end{equation}
where $\langle x_1 \ x_2 \rangle_{\rm BTZ}$ represents the bulk FPI
propagator on the spinless BTZ hole.  Here, we take $x_1$ and $x_2$ to
lie outside the geon horizon so that we may naturally associate them
with two points in an asymptotic region of the BTZ black hole.  The
first term ($\langle x_1 \ x_2 \rangle_{\rm BTZ}$) was calculated in
the geodesic approximation in section \ref{spin0BTZ} while the second
term ($\langle x_1 \ J(x_2) \rangle_{\rm BTZ}$) is given in the
geodesic approximation by (\ref{L-expr-geon}).  The geodesics that
contribute to this second term are longer than the shortest geodesic
contributing to $\langle x_1 \ x_2 \rangle_{\rm BTZ}$, so that one
might at first think that $\langle x_1 \ J(x_2) \rangle_{\rm BTZ}$ can
be neglected.  However, let us now Fourier transform this result in
order to compute the two-particle correlations in the geon state.
Since the energies of the two particles cannot add to zero, the time
translation invariance of the BTZ hole is enough to guarantee that the
contributions from the first term (with both points in the same
asymptotic region) vanish.  However, the contribution of the second
term need not vanish, corresponding to the fact that the geon does not
itself have a time translation invariance.  Thus, we see that the
two-particle correlations in the geon state can be directly tied to
the second term above, which results only from geodesics that pass through
the interior of the BTZ black hole.  In terms of the geon spacetime, 
the result is again that only geodesics passing behind the horizon
can account for the two-particle correlations.

Thus, we might try to match these correlations to those computed
in~\cite{louko:geon} for a toy model of the CFT state $|{\rm
  geon}\rangle$ dual to the ${\mathbb R}{\mathbb P}^2$ geon,
presumably with linearized quantum fluctuations in the geon vacuum
state.  The toy model replaced the CFT by a 
free scalar
field and found, in the case of a nontwisted field, the
correlations
\begin{equation} 
\label{geon2pt-jl}
\langle {\rm geon}| 
d_{n,\epsilon} d_{n',\epsilon'} 
|{\rm geon} \rangle =
\frac{
{(-1)}^n \delta_{n,n'} \delta_{\epsilon,-\epsilon'}}
 {2\sinh(\pi n/r_+)}
\ \ , 
\end{equation}
where $d_{n,\epsilon}$ is the annihilation operator for the
mode with frequency quantum number $n$, $n = 1,2,\dots$, and the
index $\epsilon$ takes the value $1$ for right-movers and $-1$ for
left-movers
(see (\ref{oscmodes}) below). 
Here and below we ignore issues involving the zero mode ($n=0$).
As a consequence of rotational invariance, the
correlations are between a right-mover and a left-mover with the same
frequency.  We note that this nontwisted free scalar field has
conformal weight $\Delta =0$.

We now show that this result can be obtained from the geodesic
approximation (\ref{L-expr-geon}) to the bulk Green's function.  It is
clear, however, that due to the simplified nature of the toy model,
one should not expect to be able to Fourier transform the asymptotic
values of the propagator and obtain (\ref{geon2pt-jl}) directly.  In
particular, the bulk propagator will not be built from only the
discrete mode spectrum of (\ref{geon2pt-jl}).  In the full interacting
CFT, the correlator will similarly not be periodic in time, and
so will not be a simple combination of these
discrete modes.  However, we may attempt to extract information
analogous to (\ref{geon2pt-jl}) by modifying the Fourier transform of
(\ref{L-expr-geon}) to take into account the fact that the bulk
propagator is not periodic in time.  We shall see that the agreement
is impressive.

In the toy (free) CFT, the oscillator modes that correspond to the
annihilation operators 
$d_{n,\epsilon}$
in (\ref{geon2pt-jl}) are
\cite{louko:geon}
\begin{equation}
u_{n,\epsilon} = \frac{1}{\sqrt{4\pi n}} 
\, e^{-in(t-\epsilon\phi)}
\label{oscmodes}
\end{equation}
where $n = 1,2,\dots$ and $\epsilon=\pm1$. 
If the Green's function 
$G(t_1,\phi_1; t_2,\phi_2)$ had
the periodicity of the oscillator modes, we would thus have 
\begin{eqnarray} 
\label{aaa}
\langle 
d_{n,\epsilon} d_{n',\epsilon'} 
\rangle 
&=&
\frac{\sqrt{n n'}}{4\pi^3} 
\int_{0}^{2\pi} dt_1 
\int_{0}^{2\pi} dt_2
\int_{0}^{2\pi} d\phi_1 
\int_{0}^{2\pi} d\phi_2 
\nonumber
\\
&&
\phantom{xxxx}
\times 
\exp[i(nt_1 + n't_2 - n\epsilon\phi_1 - n'\epsilon'\phi_2)]
\, 
G(t_1,\phi_1; t_2,\phi_2). 
\end{eqnarray}
We shall modify 
(\ref{aaa}) to take into account the lack of periodicity shortly. 

As discussed above, the part of $G(t_1,\phi_1; t_2,\phi_2)$ coming
from the geodesics that do not pass through the geon does not
contribute to $\langle d_{n,\epsilon} d_{n',\epsilon'} \rangle$.  The
part of $G(t_1,\phi_1; t_2,\phi_2)$ coming from the geodesics that do
pass through the geon is, from~(\ref{L-expr-geon}),\footnote{The
asymptotic metric in the geon spacetime is the same as in the BTZ
spacetime, so the rescaling relating the bulk and boundary propagators
is the same as in the previous subsection.}

\begin{equation}
\left(\frac{r_+^2}{2}\right)^\Delta 
\sum_{k = -\infty}^{\infty} 
\frac{1}
{ \left\{
\cosh \left[r_+(\phi_2 - \phi_1 + \pi + 2\pi k ) \right] 
+ \cosh\left[r_+ (t_2 + t_1) \right]
\vphantom{A^A_A}
\right\}^\Delta
} 
\ \ . 
\label{sum-factor}
\end{equation}
As (\ref{sum-factor}) depends on $\phi_1$ and $\phi_2$ only through
the combination 
$\phi_2 - \phi_1$, 
integrating over $\phi_2 + \phi_1$ in (\ref{aaa})
is immediate. Next, we observe that each term in 
(\ref{sum-factor}) depends on 
$\phi_2-\phi_1$ and $k$ 
only through the combination $\phi_2 - \phi_1 + 2
\pi k$.  Integrating
$\phi_2 - \phi_1$ from zero to $2\pi$ and summing over $k$ 
is thus equivalent to
integrating any one term in 
(\ref{sum-factor}) in $\phi_2 - \phi_1$ from negative infinity to positive
infinity.   Writing the integration in terms of 
the variable $y := \epsilon ( \phi_2 - \phi_1 + \pi)$, we obtain 
\begin{eqnarray} 
\label{div-t-int}
&&\langle 
d_{n,\epsilon} d_{n',\epsilon'} 
\rangle
=
\frac{\sqrt{n n'}}{2\pi^2} 
\, 
\delta_{n\epsilon, -n'\epsilon'} (-1)^n
\left(\frac{r_+^2}{2}\right)^\Delta 
\int_{0}^{2\pi} dt_1 
\int_{0}^{2\pi} dt_2
\nonumber
\\
&&
\phantom{xxxxxxxxxxxx}
\times 
\int_{-\infty}^{\infty} dy 
\, 
\frac{ \exp[i(nt_1 + n't_2 + ny)] } 
{ \left\{
\cosh \left(r_+y \right) 
+ \cosh\left[r_+ (t_2 + t_1) \right]
\vphantom{A^A_A}
\right\}^\Delta
}.
\end{eqnarray}

We must now face the fact that the integrand in (\ref{div-t-int}) is not
periodic in $t_1$ and~$t_2$.  We reinterpret (\ref{div-t-int}) by hand
so that $t_2 + t_1 := \alpha$ is integrated over
$\mathbb R$ but $t_2 - t_1$ over~$4\pi$. The integral over $t_2 - t_1$,
combined 
with the Jacobian that arises from the change of variables, yields 
then $2\pi\delta_{n,n'}$. The factor 
$\delta_{n\epsilon,
-n'\epsilon'}$ 
can thus be replaced by $\delta_{\epsilon, -\epsilon'}$,
and we find
\begin{eqnarray} 
\langle 
d_{n,\epsilon} d_{n',\epsilon'} 
\rangle
&=&
\frac{n}{\pi} 
\, 
\delta_{n, n'} 
\delta_{\epsilon, -\epsilon'} 
{(-1)}^{n}
\left(\frac{r_+^2}{2}\right)^\Delta 
\nonumber
\\
&&
\phantom{xx}
\times 
\int_{-\infty}^{\infty} d\alpha 
\int_{-\infty}^{\infty} dy 
\, 
\frac{ \exp[in(\alpha + y)] } 
{ \left[
\cosh\left(r_+\alpha \right)
+ 
\cosh \left(r_+y \right) 
\vphantom{A^A_A}
\right]^\Delta
}.
\end{eqnarray}
Changing 
variables to $u = \alpha - y$, $v = \alpha + y$, gives 
finally \cite{Grad-Rhyz-cosh} 
\begin{eqnarray} 
&&\langle 
d_{n,\epsilon} d_{n',\epsilon'} 
\rangle
= 
\frac{n}{2\pi} 
\, 
\delta_{n, n'} 
\delta_{\epsilon, -\epsilon'} 
{(-1)}^{n}
\left(\frac{r_+}{2}\right)^{2\Delta} 
\int_{-\infty}^{\infty}
\frac{ du } 
{ \left[
\cosh\left(r_+u/2\right)
\vphantom{A^A_A}
\right]^\Delta
} 
\int_{-\infty}^{\infty}
\frac{ dv \, \exp(inv) } 
{ \left[
\cosh\left(r_+v/2\right)
\vphantom{A^A_A}
\right]^\Delta
} 
\nonumber
\\
&&
\phantom{xx}
= 
\frac{n}{2\pi} 
\, 
\delta_{n, n'} 
\delta_{\epsilon, -\epsilon'} 
{(-1)}^{n}
{(r_+)}^{2(\Delta-1)} 
\left( 
\frac{\Gamma(\Delta/2)}{\Gamma(\Delta)}
\right)^2 
\, 
\Gamma\left( 
\frac{\Delta}{2} + \frac{in}{r_+}
\right) 
\, 
\Gamma\left( 
\frac{\Delta}{2} - \frac{in}{r_+}
\right). 
\label{cosh-int}
\end{eqnarray}
In the limit $\Delta \to 0_+$, (\ref{cosh-int}) reduces to 
\begin{eqnarray} 
\langle 
d_{n,\epsilon} d_{n',\epsilon'} 
\rangle 
&=& 
\frac{2n}{\pi}  
\, 
\delta_{n, n'} 
\delta_{\epsilon, -\epsilon'} 
{(-1)}^{n}
{(r_+)}^{-2} 
\, 
\Gamma\left( 
\frac{in}{r_+}
\right) 
\, 
\Gamma\left( 
- \frac{in}{r_+}
\right) 
\nonumber
\\
&=&  
\frac{4}{r_+}
\times 
\frac{
{(-1)}^n \delta_{n,n'} \delta_{\epsilon,-\epsilon'}}
{2\sinh(\pi n/r_+)}
\ \ ,
\end{eqnarray}
which agrees with (\ref{geon2pt-jl}) up to the
factor~$4/r_+$. This factor may be a consequence of 
our having neglected any pre-exponential factors in the bulk Green's
function, or from our by-hand reinterpretation of the $dt_1 \, dt_2$
integrals in~(\ref{aaa}). 

This result 
verifies the importance of geodesics passing behind the horizon
in obtaining the proper 2-particle correlations, and shows that the
toy free CFT does indeed match well with the bulk spacetime results.
As discussed earlier, it is only in special spacetimes which are
appropriately analytic than we can expect the geodesic approximation
to hold.  As a result, the fact that our calculation relies on
geodesics passing behind the horizon of the black hole is consistent
with the causal nature of the FPI propagator and with the ideas of
\cite{joeetal} that one must look beyond simple products of local
operators in the CFT to encode useful information about the interior
of a black hole.

\bigskip
\bigskip
\centerline{\bf Acknowledgements}
\medskip
 
We have enjoyed discussions with Vijay Balasubramanian, Sumit Das, Ted
Jacobson, Bernard Kay, Per Kraus, Finn Larsen, Emil Martinec, Rafael
Sorkin, Lenny Susskind, Sandip Trivedi, Bob Wald, and others at the
Val Morin workshop on Black Holes (June, 1999) and the ICTP conference
on black hole physics (July, 1999).  The work of S.F.R. was partially
supported by NSF grant PHY95-07065.  D.M. is an Alfred P. Sloan
Research Fellow and was supported in part by funds from Syracuse
University, and from NSF grant PHY-9722362.

\appendix
\section{The propagator for the rotating BTZ hole}
\label{app:spinning}

In this appendix we generalize the treatment of sections
\ref{spin0BTZ}, \ref{prop-in-BTZ} to show that the bulk geodesic
approximation method of \cite{holopart} reproduces the Green's
function in the Poincar\'e vacuum (see \cite{esko1} and the references
therein) on a single boundary component of the rotating nonextremal
BTZ hole.

The generalization of equations 
(\ref{nonrot-BTZcoords}) to the rotating case
is the rotating exterior BTZ coordinate transformation
\cite{BTZ} 
\begin{eqnarray}
&&T^1 = \sqrt{\alpha} \, \cosh(r_+ \phi - r_- t), 
\nonumber
\\
&&X^1 = \sqrt{\alpha} \, \sinh(r_+ \phi - r_- t), 
\nonumber
\\
&&T^2 = \sqrt{\alpha-1}
\, 
\sinh(r_+ t - r_- \phi), 
\nonumber
\\
&&X^2 = \sqrt{\alpha-1} 
\, 
\cosh(r_+ t - r_- \phi), 
\end{eqnarray}
with 
\begin{equation}
\alpha = 
\frac{r^2 - r_-^2}{r_+^2 - r_-^2}
\ \ ,
\end{equation}
where $r>r_+$, 
$-\infty < t < \infty$, 
and 
$-\infty < \phi < \infty$, 
and the parameters $r_\pm$ satisfy 
$0 \le r_- < r_+$. For $r_-=0$, this transformation reduces to the
spinless transformation~(\ref{nonrot-BTZcoords}). 

Introducing the points 
$y_1$ and $y_2$ in AdS$_3$ as in section~\ref{spin0BTZ},  
respectively at 
$(t_1,r,\phi_1)$ and
$(t_2,r,\phi_2)$, we find 
\begin{eqnarray}
D(y_1,y_2) 
&=& 
2\alpha
\left[ \cosh(r_+\Delta \phi- r_- \Delta t) -1 \right] 
\nonumber
\\
&& 
- 
2 (\alpha -1)
\left[ \cosh(r_+ \Delta t - r_- \Delta \phi ) -1 \right]
\ \ . 
\label{D-in-sBTZ}
\end{eqnarray}
When $t_1$, $t_2$, $\phi_1$, and $\phi_2$ are fixed, 
and such that $|\Delta \phi| > |\Delta t|$, 
equation 
(\ref{D-in-sBTZ}) shows that $D(y_1,y_2)>0$ for sufficiently
large~$r$. $y_1$~and $y_2$ can then be joined by a spacelike
geodesic, and the length 
$L(y_1,y_2)$ of this geodesic has the large $r$ expansion
\begin{equation}
\exp\left[L(y_1,y_2)\right]
= 
\frac{2 r^2}{r_+^2} 
\left[
\cosh(r_+\Delta \phi - r_- \Delta t) 
- \cosh(r_+\Delta t - r_- \Delta \phi ) 
\right]
+ O\left( 1 \right) 
\ \ . 
\label{L-expr-spin}
\end{equation}

Now, in 
the region of AdS$_3$ 
covered by the exterior BTZ coordinates, the rotating
BTZ quotient construction amounts to the identification $(t,r,\phi)
\sim (t,r,\phi+2\pi)$.  Let again $x_1$ (respectively $x_2$) be the
equivalence class of the point $y_1$~($y_2$). Assuming $|\Delta \phi +
2\pi n| > |\Delta t|$ for all integers~$n$, and proceeding as in
section~\ref{prop-in-BTZ}, we find that the geodesic approximation to
the path integral (\ref{picor}) reads
\begin{eqnarray}
&&\langle \phi (x_1) \phi(x_2) \rangle_{\rm FPI} 
= \int d{\cal P} e^{i
\Delta L({\cal P})}
\approx 
\sum_n\exp\left[-\Delta L_n(x_1,x_2)\right]
\nonumber
\\
&&=
\left(\frac{r_+^2}{2 r^2} \right)^\Delta
\sum_{n=-\infty}^{\infty} 
\frac{1}
{
\left\{
\cosh\left[r_+(\Delta \phi+ 2\pi n ) - r_-\Delta t \right] - 
\cosh\left[r_+\Delta t - r_- (\Delta \phi + 2\pi n ) \right]
\vphantom{A^A_A}
\right\}^\Delta
}
\nonumber
\\
&&
\phantom{xx}
+ 
O\left( \left( \frac{r_+^2}
{r^2} \right)^{\Delta+1}
\right) 
\ \ . 
\label{BTZspin-geodapp-expansion} 
\end{eqnarray}
The boundary-dependent factor in the leading term in 
(\ref{BTZspin-geodapp-expansion}) at $r\to\infty$ can be rewritten as 
\begin{equation}
\sum_{n=-\infty}^{\infty} 
\frac{1}
{
\left\{
\sinh
\left[ 
\frac{1}{2} 
(r_+ - r_-)(\Delta \phi + \Delta t + 2\pi n)
\right]
\sinh
\left[ 
\frac{1}{2}
(r_+ + r_-)(\Delta \phi - \Delta t + 2\pi n)
\right]
\right\}^\Delta
}
\ \ , 
\end{equation}
which is recognized as the dominant factor in the Green's function in
the Poincar\'e vacuum on the boundary of the rotating BTZ hole (see
\cite{esko1} and the references therein).


\end{document}